\documentclass[3p,times]{elsarticle}

\usepackage{ecrc}
\graphicspath{{figures/}}

\volume{00}

\firstpage{1}

\journalname{Physics Reports}

\runauth{}

\jid{procs}

\jnltitlelogo{Physics Reports}

   \CopyrightLine{2016}{Elsevier Ltd. All rights reserved}

\usepackage{amsmath,amssymb}

\biboptions{square,comma,sort&compress}

\usepackage[figuresright]{rotating}
\usepackage{booktabs}
\usepackage{tabularx}

\usepackage{multirow}

\usepackage{wrapfig}
\usepackage{placeins}
\usepackage[colorlinks=true,allcolors=blue]{hyperref}

\begin{document}

\begin{frontmatter}

\dochead{}

\newcommand{\sineff}{$\sin^2\theta^\ell_{\rm eff}~$}

\newcommand{\Korea}{Department of Physics and Astronomy, Seoul National University, Seoul 151-747, Korea}
\newcommand{\Rochester}{Department of Physics and Astronomy, University of Rochester, Rochester, NY  14627, USA}

\title{Precision Measurements of the Effective Leptonic Weak Mixing Angle at Hadron Colliders}

\author{Hyon-San~Seo}
\address{\Rochester}

\author{Arie~Bodek}
\address{\Rochester}

\author{Un-Ki~Yang}
\address{\Korea}

\date{Aug. 3, 2025}

\begin{abstract}
We review precision determinations of the effective leptonic weak mixing angle, $\sin^2\theta^\ell_{\rm eff}$, from Drell--Yan angular asymmetries at the Tevatron and the LHC.  As statistical uncertainties have dramatically decreased, parton distribution function (PDF) uncertainty now dominates the most precise hadron-collider results.
We describe the angular observables, electroweak calculation schemes, and advanced methods such as replica reweighting and Hessian profiling, used to constrain PDFs directly from measured angular distributions. To identify the origin of differences between modern PDF sets (e.g., CT18, MSHT20, NNPDF) in the extracted mixing angle, we employ a simplified leading-order Z-pole calculation that separates quark-flavor production weights from quark-direction dilution. We demonstrate how these PDF-induced differences can be mitigated by incorporating complementary constraints from W/Z cross-section ratios and W-boson decay asymmetries. Furthermore, we discuss the requirements for a common reinterpretation and future combination of LHC measurements, highlighting how correlated PDF uncertainties limit the ultimate combined precision. 
Finally, we outline future prospects, including complementary measurements in $Z+b$ production and high-mass tests of the running $\overline{\rm MS}$ weak mixing angle.
\end{abstract}


\end{frontmatter}

\section{Introduction}
\label{Intro}
The weak mixing angle is a fundamental cornerstone of the Standard Model (SM) of particle physics. It describes the mixing of the neutral electroweak gauge fields into the photon and the Z boson. Precision measurements of this parameter provide one of the most stringent consistency tests of the electroweak sector and serve as powerful, model-independent probes for physics beyond the Standard Model (BSM).
Because the effective mixing angle absorbs higher-order electroweak radiative corrections, it is highly sensitive to the presence of new BSM particles running in quantum loops.
For instance, theoretical extensions such as two-Higgs-doublet models or theories with an additional $U(1)$ gauge symmetry (predicting a $Z'$ boson) can shift the electroweak radiative corrections and modify the observed leptonic Z couplings~\cite{Grimus:2008nb,Langacker:2008yv}.
Therefore, pushing the precision limits of this parameter serves as a highly sensitive probe 
for these BSM scenarios. 
This sensitivity to new physics can be enhanced by combining measurements of the mixing angle with those of the W-boson mass, which depends on a different combination of electroweak radiative corrections.

At tree level, the weak mixing angle can be expressed in terms of either the $SU(2)_L$ and $U(1)_Y$ gauge couplings ($g$ and $g^\prime$), or the W- and Z-boson masses ($m_W$ and $m_Z$):
\begin{equation}
\sin^2\theta_W=\frac{g^{\prime\,2}}{g^2+g^{\prime\,2}}
=1-\frac{m_W^2}{m_Z^2}
\qquad\text{(tree level)}.
\label{eq:tree_weak_angle}
\end{equation}
Beyond tree level, radiative corrections modify the relation between these two expressions, making the definition of the mixing angle dependent on the chosen renormalization scheme. For example, 
the on-shell scheme retains the mixing angle using the physical boson masses 
($\sin^2\theta_W^{\rm OS}\equiv1-m_W^2/m_Z^2$~\cite{PDG2026EW}), 
while the $\overline{\rm MS}$ scheme introduces a scale-dependent running parameter based 
on renormalized gauge couplings. 
To avoid these scheme-dependent theoretical conversions when reporting experimental data, precision asymmetry measurements are typically reported in terms of the effective leptonic weak mixing angle,
$\sin^2\theta^\ell_{\rm eff}$.
This \emph{pseudo-observable}~\cite{ALEPH:2005ab,Dubovyk:2019ewpo} is directly related to the measurable ratio of the effective vector ($g_V^\ell$) and axial-vector ($g_A^\ell$) couplings of the Z boson to leptons at the Z pole.
\begin{equation}
\sin^2\theta^\ell_{\rm eff}
=\frac{1}{4}
\left(1-\operatorname{Re}\left[\frac{g_V^\ell}{g_A^\ell}\right]\right).
\label{eq:effective_leptonic_angle}
\end{equation}

\begin{figure}[!t]
\centering
\includegraphics[width=6.2in]{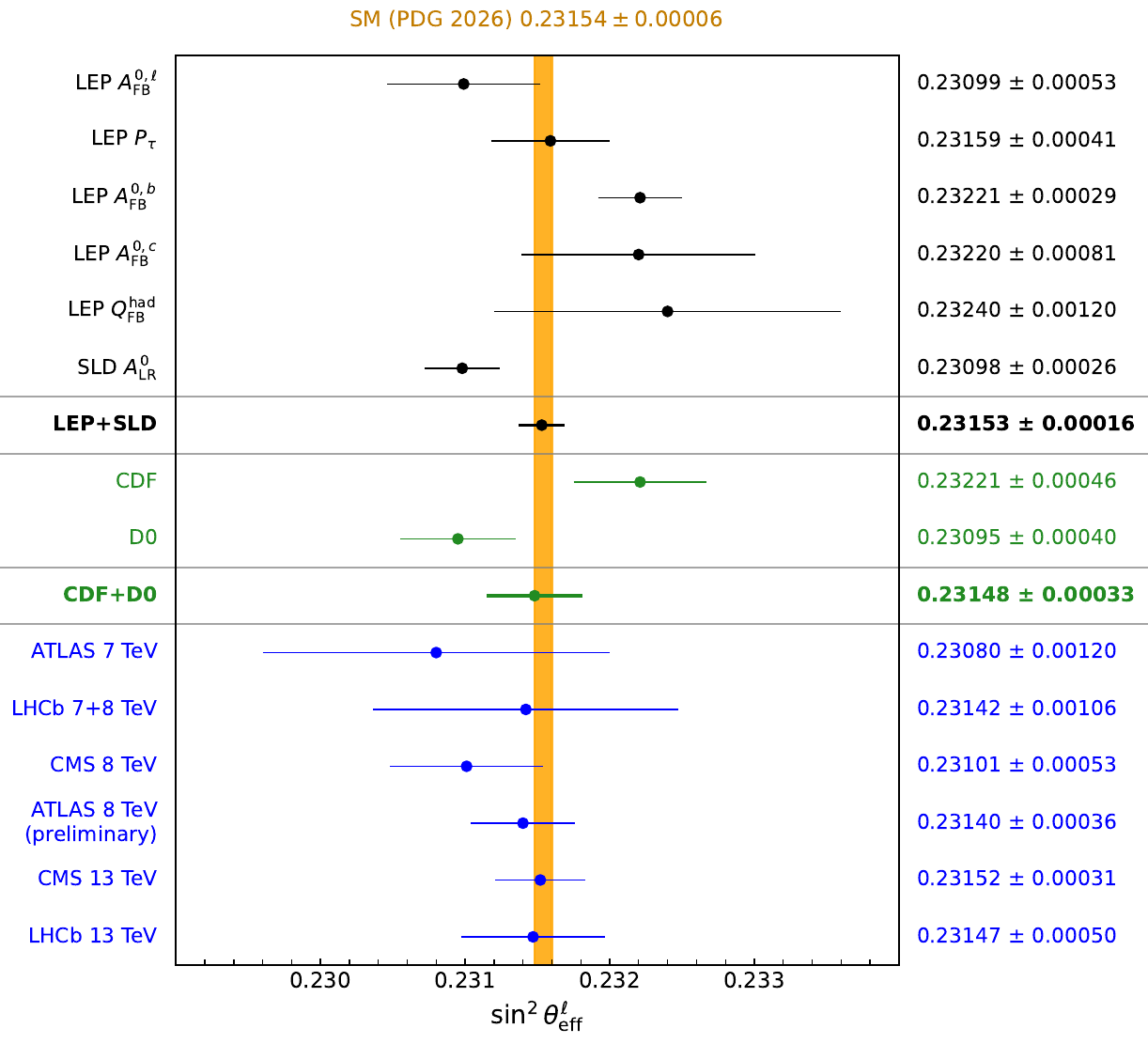}
\caption{Measurements of $\sin^2\theta^\ell_{\rm eff}$~\cite{ATLAS:2015ihy,LHCb:2015jyu,CMS:2018ktx,ATLAS:2018gqq,CMS:2024ony,LHCb:2024ygc}.  Error bars show total uncertainties.  Bold entries, separated by horizontal lines, indicate the LEP+SLD~\cite{ALEPH:2005ab} and CDF+D0~\cite{CDFD0:2018qhk} combinations.  The ATLAS 8 TeV result is preliminary.  The orange band shows the 2026 PDG SM prediction, calculated without using weak-mixing-angle measurements as inputs~\cite{PDG2026EW}.}
\label{Fig_1}
\end{figure}

Historically, the most precise determinations of $\sin^2\theta^\ell_{\rm eff}$ came from electron-positron
($e^+e^-$) collisions at the Large Electron-Positron collider (LEP) and  the SLD experiment at the Stanford Linear Collider. By measuring various parity-violating observables, including the forward-backward asymmetry, the left-right asymmetry, and the tau polarization, these experiments provided six individual determinations, which are summarized in Fig.~\ref{Fig_1}.
They achieved a combined value of 
$\sin^2\theta^\ell_{\rm eff}=0.23153\pm0.00016$. While this combined value is in excellent agreement with the SM prediction, the internal consistency of the six measurements exhibits $\chi^2/\mathrm{ndf}=11.8/5$, corresponding to a p-value of approximately $4\%$~\cite{ALEPH:2005ab}. This level of internal spread provides a motivation for independent determinations.

Today, the landscape of precision electroweak physics has expanded to hadron colliders, which offer a compelling testing ground for this purpose. 
Measurements of the angular distributions in Drell–Yan dilepton production 
($q\bar q\to\gamma^*/Z\to\ell^+\ell^-$) at the Tevatron and the LHC (also shown in Fig.~\ref{Fig_1}) utilize completely different initial states and involve experimental systematics that are largely orthogonal to those of $e^+e^-$
colliders. Consequently, precision determinations at hadron colliders are not merely complementary; they are crucial for providing an independent perspective on this fundamental parameter.

However, extracting precision electroweak parameters at hadron colliders presents a unique and profound challenge: the initial state involves composite protons (or antiprotons). The predicted angular distributions heavily depend on the parton distribution functions (PDFs) of the incoming hadrons. Because the precise momentum fraction and direction of the interacting quark and antiquark are not known on an event-by-event basis, incorrect quark direction assignments dilute the measured forward-backward asymmetry. Furthermore, the relative contributions from different quark flavors (e.g., up, down, strange) strongly depend on the PDFs. Consequently, as the statistical uncertainties have been substantially reduced
with the large LHC datasets, the uncertainty associated with PDFs has become the dominant barrier 
in the most precise hadron-collider determinations of $\sin^2\theta^\ell_{\rm eff}$.

This review provides a comprehensive overview of precision effective weak mixing angle measurements at hadron colliders, with a central focus on the methodologies developed to understand and constrain the associated PDF uncertainties. In Section 2, we describe the relevant angular observables, electroweak calculations, and modern techniques for constraining PDFs directly from measured angular distributions, including replica reweighting and Hessian profiling. Section 3 reviews the measurements and analysis strategies from the Tevatron (CDF and D0), and LHC (ATLAS, CMS, and LHCb) experiments. In Section 4, we employ a simplified leading-order Z-pole calculation to identify the origin of PDF-set dependence, separating quark-flavor production weights from quark-direction dilution. Section 5 discusses the requirements and challenges for a common reinterpretation and future combination of LHC measurements, highlighting the impact of correlated PDF uncertainties. Finally, in Section 6, we discuss future prospects, including complementary constraints from  $Z+b$ production and high-mass tests of the running weak mixing angle above the Z pole.

\section{Angular observables and extraction methods}
\label{sec:analysis_strategy}

This section describes how Drell--Yan angular measurements are used to extract the effective leptonic weak mixing angle.  We first introduce the angular observables, quark-direction dilution, and the treatment of acceptance and detector effects.  We then discuss the electroweak corrections used in the extraction and the methods for constraining PDFs with the measured angular distributions.

The forward--backward asymmetry is measured from the polar-angle distribution of the negative lepton.  In the Collins--Soper frame~\cite{Collins:1977iv}, this angle is denoted by $\theta$, with the forward direction assigned using an estimate of the incoming quark direction.  After integration over the azimuthal angle $\phi$, the distribution takes the form
\begin{equation}
\frac{d\sigma}{d\cos\theta}\propto
1+\cos^2\theta +  \frac{A_0}{2} (1-3\cos^2\theta)  +A_4\cos\theta.
\label{eq:hadronic_angular_distribution}
\end{equation}
The term $A_4\cos\theta$ is odd under $\cos\theta\to-\cos\theta$ and determines the forward--backward difference.  For full angular acceptance, integration over $\cos{\theta}$ yields $A_{\rm FB}^{\rm full}=(3/8)A_4$.  The term proportional to $A_0$ integrates to zero for full angular acceptance but can affect the asymmetry normalization after lepton cuts.  Acceptance corrections therefore depend on the QCD modeling of the dilepton recoil and angular coefficients.  The angular  coefficients $A_4$ and $A_0$ depend on the dilepton invariant mass $m\equiv m_{\ell\ell}$, rapidity $y\equiv y_{\ell\ell}$, and transverse momentum.  The angular coefficient $A_4(m,y)$  (integrated over transverse momentum) is used in the extraction of the weak mixing angle.   Near the Z pole, the dependence of $A_4$ on the  leptonic Z couplings provides sensitivity to $\sin^2\theta^\ell_{\rm eff}$.

To predict $A_4$ or $A_{\rm FB}$, the partonic angular distributions must be combined with the PDFs of the incoming hadrons.  At leading order in QCD, the contributions from $q\bar q\to\gamma^*/Z\to\ell^+\ell^-$ have the form
\begin{equation}
\sum_q\left[
f_{q/h_1}(x_1,\mu_F^2)f_{\bar q/h_2}(x_2,\mu_F^2)\hat\sigma_{q\bar q}
 + f_{\bar q/h_1}(x_1,\mu_F^2)f_{q/h_2}(x_2,\mu_F^2)\hat\sigma_{\bar q q}
\right],
\end{equation}
Here $h_1$ and $h_2$ denote the incoming hadrons, and $\hat\sigma$ denotes the partonic cross section.  The momentum fractions are $x_{1,2}=(m/\sqrt{s})e^{\pm y}$, with factorization scale $\mu_F\simeq m$.  The quark flavors contribute with different electroweak couplings and partonic asymmetries and their PDF weights affect the predicted $A_{\rm FB}$.  The two terms correspond to the quark entering from opposite beams, whose relative contributions also affect the quark-direction assignment.  Higher-order QCD predictions include radiation and additional channels such as the quark-gluon ($qg$) process.  The displayed expression illustrates the flavor and direction effects at Born level.

In proton--antiproton collisions at the Tevatron, the incoming quark is more likely to come from the proton than from the antiproton.  The proton contains valence quarks, while the antiproton contains valence antiquarks, favoring this assignment.  The proton-beam direction is therefore used to define the forward direction.  Events with an antiquark from the proton and a quark from the antiproton are assigned the opposite quark direction.  Their contribution to the forward--backward difference changes sign and dilutes the asymmetry.

In proton--proton collisions at the LHC, the two beams have the same parton distributions, so neither beam is a preferred quark source.  However, valence quarks typically carry a larger momentum fraction than sea antiquarks.  The longitudinal boost direction of the dilepton system is therefore used to estimate the quark direction and define the forward direction.  At small $|y|$, the incoming partons have similar momentum fractions.  The boost direction then gives a less reliable estimate of the quark direction, and the asymmetry is strongly diluted.  At larger $|y|$, the momentum imbalance generally improves the assignment and  the dilution is reduced.  Both the dilution and the flavor weights depend on PDFs and vary with dilepton mass and rapidity.  Measurements in mass and rapidity bins therefore provide information on PDFs as well as sensitivity to $\sin^2\theta^\ell_{\rm eff}$.

A forward--backward asymmetry can be measured as $(N_F-N_B)/(N_F+N_B)$, where $N_F$ and $N_B$ are the numbers of events with positive and negative $\cos\theta$, respectively.  This uses only the sign of $\cos\theta$, although the sensitivity to $A_4$ varies with $|\cos\theta|$.  Fitting the angular distribution or comparing yields in matched angular bins retains this additional information.  At fixed dilepton kinematics, the local asymmetry in narrow bins at $+c$ and $-c$, with $c=|\cos\theta|$, follows
\begin{equation}
\frac{N(c)-N(-c)}{N(c)+N(-c)}
=\frac{A_4 c}{1+c^2+\tfrac{A_0}{2}(1-3c^2)}.
\label{eq:local_angular_asymmetry}
\end{equation}
Here $N$ denotes the signal yield.  Acceptance and efficiency may vary substantially with $c$, while remaining similar between the matched positive and negative bins. In the limit of equal responses and negligible bin migration, their common factor cancels between the numerator and denominator.  Residual forward--backward differences are often much smaller and can be included, together with bin migration effects, in reconstructed-level templates or unfolding.

The angular dependence can also be used without explicit binning in $\cos\theta$.  Angular weighting uses the $c$ dependence in Eq.~\ref{eq:local_angular_asymmetry} to combine information on the common coefficient $A_4$ from different angles, with weights reflecting the statistical sensitivity at each angle~\cite{Bodek:2010qg}.  The weighted event sums give a single asymmetry estimate, $A_{\rm FB}^{w}$, in each dilepton mass and rapidity bin.  In the ideal limit discussed above, this estimates $A_{\rm FB}^{\rm full}=(3/8)A_4$.  The method can improve statistical sensitivity relative to simple counting.  Avoiding an additional angular binning also reduces the number of analysis bins and the need to handle sparsely populated bins.

The measured $A_{\rm FB}$ or $A_4$ distributions provide the input to a fit of $\sin^2\theta^\ell_{\rm eff}$.  This fit can be performed at reconstructed level using templates that include detector effects and QED final-state radiation (FSR), or after unfolding to a specified lepton definition.  For unfolded results, the acceptance and treatment of FSR define the quantity being fitted.  In either approach, the mixing angle is varied in the theory prediction, whose electroweak and PDF dependence is discussed below~\cite{CMS:2018ktx,CMS:2024ony}.

\subsection{Electroweak corrections and input schemes}
\label{sec:ew_input_schemes}

The effective leptonic mixing angle is defined by the leptonic Z-pole coupling ratio in Eq.~\ref{eq:effective_leptonic_angle}.  In hadron-collider measurements, the fitted angular distributions also depend on quark couplings and cover a range of dilepton masses.  An electroweak prescription is therefore needed to relate the fitted mixing angle to these distributions, including flavor-dependent weak corrections, photon--Z interference, and the mass dependence of the form factors~\cite{ALEPH:2005ab,Chiesa:2019nqb}.

Earlier extractions incorporated weak corrections through effective couplings or electroweak form factors while retaining the Born structure of the neutral-current amplitude.  As experimental precision improved, more complete implementations accounted for differences between the effective couplings of leptons and quarks, complex and mass-dependent form factors, and photon vacuum polarization~\cite{CDFD0:2018qhk}.  Such form-factor treatments are commonly described as improved or enhanced Born approximations (IBA/EBA).  The form factors can be calculated with \textsc{ZFITTER}, which uses the \textsc{Dizet} library for the weak corrections~\cite{Arbuzov:2005ZFITTER,RichterWas:2019TauSpinnerEW}.

A further step was to define the fitted effective mixing angle as an independent renormalized input of the electroweak calculation.  The $(G_\mu,m_Z,\sin^2\theta^\ell_{\rm eff})$ scheme was formulated for Drell--Yan calculations in 2019~\cite{Chiesa:2019nqb}.  In this scheme, the mixing angle can be varied directly while the quark couplings and off-pole corrections are calculated consistently.  The renormalization conditions preserve the leptonic pole definition at each perturbative order.  This scheme is implemented in \textsc{Z\_ew-BMNNPV} within \textsc{POWHEG-BOX-V2}, allowing the effective mixing angle to be fitted directly in a full NLO electroweak calculation of Drell--Yan production~\cite{Barze:2013fru,Chiesa:2019nqb,Chiesa:2024qzd}.

Using $\sin^2\theta^\ell_{\rm eff}$ as an input absorbs important pole corrections into its definition.  Near the Z pole, published comparisons find smaller residual universal higher-order corrections and a weaker top-quark-mass dependence in $A_{\rm FB}$ than in other widely used input schemes, such as $(G_\mu,m_W,m_Z)$ and $(\alpha(0),G_\mu,m_Z)$~\cite{Chiesa:2019nqb,Chiesa:2024qzd}.  These features make the $\sin^2\theta^\ell_{\rm eff}$ input scheme well suited for precision mixing angle extractions, and it has been adopted in the recent CMS and LHCb measurements using 13 TeV Run 2 data samples~\cite{CMS:2024ony,LHCb:2024ygc}.

\subsection{PDF constraints from angular asymmetries}
\label{sec:pdf_constraints}

PDFs are a leading source of systematic uncertainty in hadron-collider determinations of $\sin^2\theta^\ell_{\rm eff}$.  Their relative importance increases as larger data samples reduce the statistical uncertainty.  Different PDF sets may yield different extracted mixing angles from the same measured asymmetry  distributions~\cite{CMS:2024ony}.  Therefore, both the uncertainty within a PDF set and the dependence of the extracted  mixing angle on the choice of PDF set need to be understood.

\begin{figure*}[t]
\centering
\includegraphics[width=\textwidth]{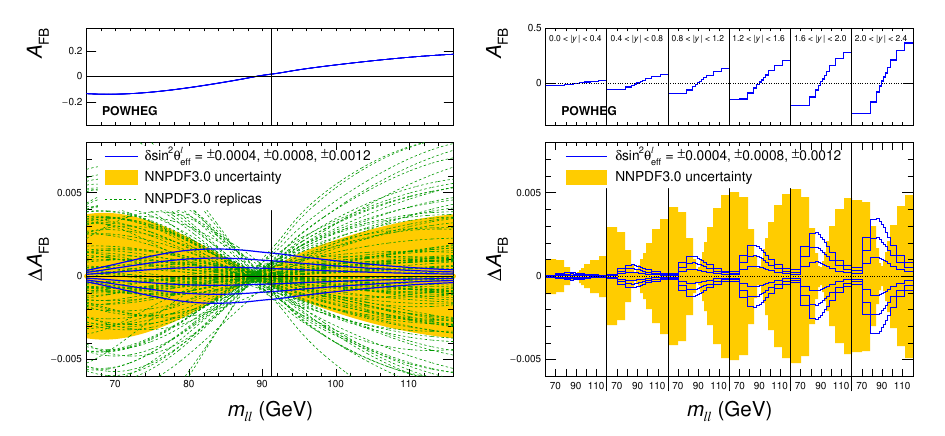}
\caption{POWHEG predictions for the dilepton mass dependence of $A_{\rm FB}$ in the CMS 8 TeV analysis, taken from Fig.~5 of Ref.~\cite{CMS:2018ktx}. The left panels integrate over rapidity, while the right panels show six rapidity bins. Upper panels give the predictions at $\sin^2\theta^\ell_{\rm eff}=0.23120$. Lower panels show changes in $A_{\rm FB}$ from mixing angle variations (solid curves) and 100 NNPDF3.0 replicas (dashed curves). The shaded bands indicate the PDF standard deviation.}
\label{Fig_afb_angle_pdf_shapes}
\end{figure*}

The dilepton mass and rapidity dependence of the measured $A_{\rm FB}$ or $A_4$ distributions can be used to constrain the PDFs within the same fit that is used in the extraction of  $\sin^2\theta^\ell_{\rm eff}$~\cite{Bodek:2015ljm}.  Changing the mixing angle mainly affects the leptonic Z couplings, while PDF variations change the quark-flavor mixture and quark-direction dilution.  These effects have different mass and rapidity dependences, as illustrated in Fig.~\ref{Fig_afb_angle_pdf_shapes}.  A fit in both mass and rapidity bins can therefore distinguish some PDF variations from a mixing angle shift.  This can reduce both the propagated PDF uncertainty and the spread in fitted mixing angles extracted with different PDF sets.

\suppressfloats[t]

PDF uncertainties are commonly provided as Monte Carlo replicas or Hessian eigenvector variations~\cite{NNPDF:2021njg,Hou:2019efy,Bailey:2020ooq}.  In both representations, the published error ensemble supplies the prior PDF uncertainty, while the measured asymmetries enter through a common data likelihood fit.  The mixing angle is fitted together with the PDF variations (profiling).  Let $s_{\rm eff}^2=\sin^2\theta^\ell_{\rm eff}$, let $\boldsymbol d$ contain the measured asymmetries, and let $\boldsymbol t(s_{\rm eff}^2,f)$ denote the predictions for a PDF $f$.  The data contribution is
\begin{equation}
\chi_A^2(s_{\rm eff}^2,f)=
[\boldsymbol d-\boldsymbol t(s_{\rm eff}^2,f)]^{\mathsf T}
V^{-1}
[\boldsymbol d-\boldsymbol t(s_{\rm eff}^2,f)],
\label{eq:asymmetry_pdf_likelihood}
\end{equation}
where $V$ is the covariance matrix of the measured distributions.  Systematic effects can instead be treated with constrained nuisance parameters in the fit.

In the replica representation, the PDF prior is sampled by an ensemble of equally weighted members $f_r$.  For each member, one fits $s_{\rm eff}^2$, obtaining a best-fit value $s_{{\rm eff},r}^2$ and a minimum $\chi^2_{A,r}$.  The Giele--Keller (GK) prescription assigns normalized likelihood weights~\cite{Giele:1998reweight,Sato:2014reweight}
\begin{equation}
\begin{aligned}
w_r&=\frac{\exp(-\chi^2_{A,r}/2)}
{\sum_k\exp(-\chi^2_{A,k}/2)},\qquad
\overline{s_{\rm eff}^2}=\sum_r w_r s_{{\rm eff},r}^2,\\
\sigma_{\rm PDF}^2&=\sum_r w_r(s_{{\rm eff},r}^2-\overline{s_{\rm eff}^2})^2.
\end{aligned}
\label{eq:gk_pdf_weights}
\end{equation}
Replicas that give a poorer description of the data receive smaller weights.  The weighted spread of $s_{{\rm eff},r}^2$ in Eq.~\ref{eq:gk_pdf_weights} represents the PDF component.  The within-member fit uncertainty is also retained when determining the total uncertainty.  The effective ensemble size, $N_{\rm eff}=1/\sum_r w_r^2$ in this normalization, measures whether sufficiently many replicas contribute to the result.

In the Hessian representation, PDF variations are described by eigenvector directions about the central fit.  Let $\beta_k$ be PDF nuisance parameters normalized to the published one-standard-deviation variations, with a confidence-level conversion applied where necessary.  To linear order, the prediction and the profiling objective can be written as
\begin{equation}
\begin{aligned}
\boldsymbol t(s_{\rm eff}^2,\boldsymbol\beta)
&\simeq \boldsymbol t_0(s_{\rm eff}^2)+\sum_k\boldsymbol D_k(s_{\rm eff}^2)\,\beta_k,\\
\chi_{\rm prof}^2(s_{\rm eff}^2,\boldsymbol\beta)
&=\chi_A^2(s_{\rm eff}^2,\boldsymbol\beta)+\sum_k\beta_k^2,
\end{aligned}
\label{eq:hessian_pdf_profile}
\end{equation}
where $\boldsymbol D_k$ is obtained from the PDF error members, for example from half their positive-minus-negative difference for a symmetric eigenvector pair.  The quadratic penalty encodes the Gaussian PDF prior.  Minimization shifts and constrains the PDF directions favored by the data.  At each trial $s_{\rm eff}^2$, the profiled PDF and other nuisance parameters are minimized, retaining their correlations with the mixing angle.  A unit increase of the profiled objective gives the usual one-parameter interval in the Gaussian approximation.  For linear predictions, a Gaussian prior, and a common data likelihood and uncertainty normalization, GK replica reweighting and Hessian profiling give equivalent central values and covariance with adequate replica sampling~\cite{Paukkunen:2014reweight}.

Some global Hessian PDF fits use a tolerance to account for tensions among datasets and limitations of the parametrization and theory~\cite{Hou:2019efy,Bailey:2020ooq}.  Instead of the usual $\Delta\chi^2=1$ criterion, they allow a larger increase in the global-fit objective, $\Delta\chi^2=T^2$, where $T>1$ can depend on the eigenvector direction.  Profiling uses the resulting broader uncertainty as a PDF prior.  The prior constraint is therefore weaker, while the asymmetry measurement retains its measured covariance.  This treatment can give different results from a global refit including the same data.

In these precision electroweak extractions, because no additional tolerance is applied to the asymmetry data, their experimental covariance and the theory uncertainties must be reliably evaluated.  In measurements of asymmetries and cross section ratios   common normalization, efficiency and acceptance  uncertainties mostly cancel, and in statistically dominated bins the covariance is largely determined by event counts.  The fit quality, the shifts and constraints of individual nuisance parameters, and the stability of the extracted mixing angle with alternative PDF sets should be checked.  Using the profiled PDFs outside the probed $(x,\mu_F)$ region requires independent validation.

Section~\ref{sec:measurements} describes how these methods are used in individual measurements.  Table~\ref{tab:hadron_measurements} summarizes the electroweak and PDF treatments and uncertainties of selected measurements.

\section{Tevatron and LHC measurements}
\label{sec:measurements}
\suppressfloats[t]

We review the CDF and D0 measurements of $\sin^2\theta^\ell_{\rm eff}$ and their Tevatron combination, followed by the ATLAS, CMS, and LHCb results.  For each experiment, we describe the angular observables and kinematic coverage, the electroweak and PDF treatments used in the extraction, and the main sources of uncertainty.  Figure~\ref{Fig_1} compares the reported values, and Table~\ref{tab:hadron_measurements} summarizes the analysis methods and uncertainties.

\begin{figure*}[!htb]
\centering
\includegraphics[width=5.8in]{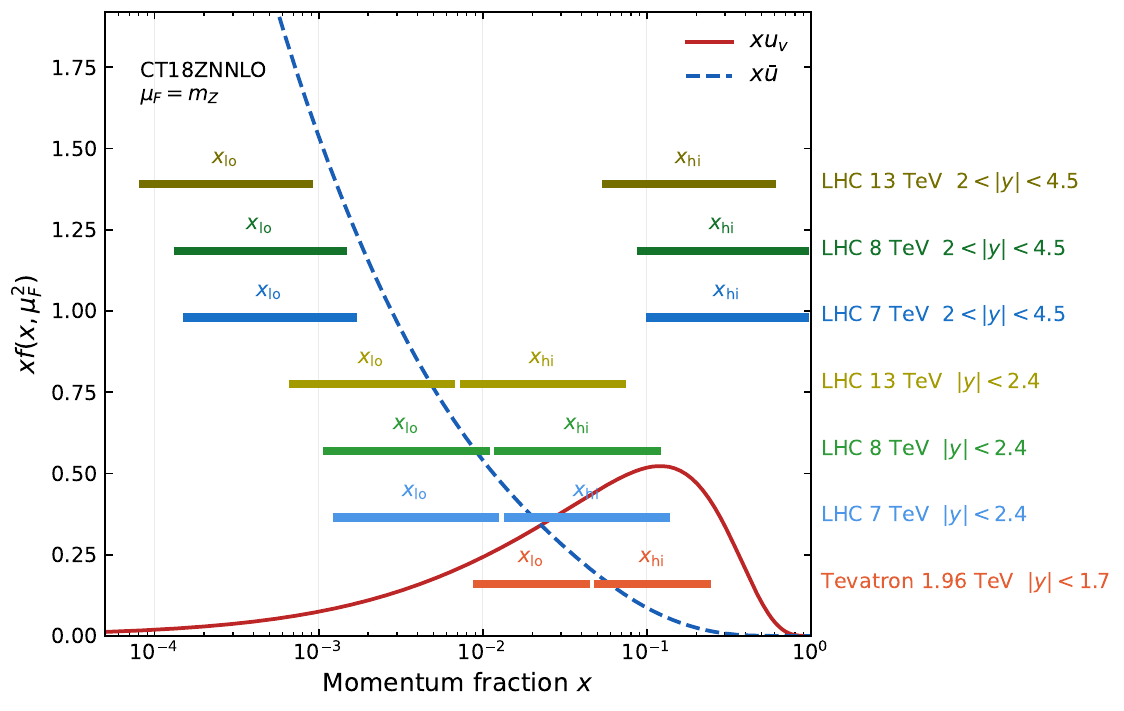}
\caption{Parton momentum-fraction ranges near the Z pole for representative Tevatron and LHC rapidity intervals.  The horizontal bars show $x_{\rm lo}$ and $x_{\rm hi}$ from the leading-order relation $x_{\rm hi,lo}=(m_Z/\sqrt{s})e^{\pm|y|}$, using the collision energies and dilepton rapidity ranges indicated.  The solid and dashed curves show CT18Z NNLO $xu_v$ and $x\bar u$, respectively, at $\mu_F=m_Z$, with $u_v=u-\bar u$.}
\label{Fig_x_ranges_lhc}
\end{figure*}

The Tevatron and LHC measurements have different strengths and limitations.  At the Tevatron, quark-direction dilution is relatively small, as discussed in Section~\ref{sec:analysis_strategy}, but the available Drell--Yan statistics limit the precision.  ATLAS and CMS benefit from larger cross sections and the high integrated luminosities accumulated at the LHC.  In their central regions, however, the asymmetry is more strongly diluted and the extraction is more sensitive to PDFs than at the Tevatron.  Forward-lepton channels extend the rapidity coverage, while binning in rapidity retains the different sensitivities of central and forward events.  LHCb benefits from smaller dilution and PDF uncertainties in the forward region, but its lower integrated luminosity and restricted acceptance leave the measurement limited by statistics~\cite{LHCb:2024ygc}.  Figure~\ref{Fig_x_ranges_lhc} compares the momentum-fraction ranges probed by these acceptances.

Tau polarization provides a complementary probe of the leptonic Z couplings without estimating the incoming quark direction.  ATLAS measured this observable at 8 TeV~\cite{ATLAS:2017taupol}, and CMS extracted $\sin^2\theta^\ell_{\rm eff}$ from its 13 TeV measurement~\cite{CMS:2024taupol}.  The following discussion focuses on electron and muon Drell--Yan asymmetries, which currently give more precise mixing-angle determinations.

\subsection{Tevatron measurements}

CDF and D0 measured the Drell--Yan forward--backward asymmetry in electron and muon final states at $\sqrt{s}=1.96$ TeV.  In proton--antiproton collisions, the incoming quark is more likely to originate from the proton, allowing the proton-beam direction to define forward.  The resulting asymmetry is less diluted than in proton--proton collisions.

CDF used approximately 9 fb$^{-1}$ of  Tevatron Run II data in each lepton channel.  Angular weighting retained information on the variation of the statistical sensitivity with the lepton angle.  The resulting $A_{\rm FB}^{w}(m)$ distributions were unfolded for detector effects and QED final-state radiation, then compared with \textsc{POWHEG} templates incorporating \textsc{ZFITTER} electroweak form factors.  NNPDF3.0 NNLO replicas were reweighted using the measured asymmetries, allowing the same data to constrain the PDFs used in the mixing angle extraction.  The combined electron--muon result was $\sin^2\theta^\ell_{\rm eff}=0.23221\pm0.00046$~\cite{CDF:2016cei,CDFD0:2018qhk}.

D0 used 9.7 fb$^{-1}$ of electron data and 8.6 fb$^{-1}$ of muon data.  It determined $A_{\rm FB}(m)$ from forward and backward event counts, integrated over the accepted rapidity range.  The mixing angle was extracted by comparison with reconstructed-level templates that included the detector response.  Periodic reversals of the magnetic fields helped suppress detector-induced charge asymmetries~\cite{D0:2017ekd,CDFD0:2018qhk}.

The final CDF and D0 central values were expressed using a common \textsc{ZFITTER}-based electroweak prescription and NNPDF3.0 NNLO PDFs.  For D0, this involved corrections to the mixing angles extracted with \textsc{Pythia} templates, including the update from NNPDF2.3 to NNPDF3.0 for the electron channel.  The PDF uncertainties retained the original replica evaluations, without reweighting to the D0 asymmetry data.  The corrected electron and muon results gave $\sin^2\theta^\ell_{\rm eff}=0.23095\pm0.00040$~\cite{D0:2017ekd,CDFD0:2018qhk}.

The CDF and D0 results were then combined with BLUE, treating their PDF uncertainties as fully correlated.  The Tevatron result was
\begin{equation}
\sin^2\theta^\ell_{\rm eff}=0.23148\pm0.00033
\end{equation}
with statistical and PDF contributions of $0.00027$ and $0.00018$, respectively.  The combined precision remained mainly limited by statistics.  The compatibility test of the CDF and D0 inputs gives a p-value of $2.6\%$, providing additional motivation for independent measurements at the LHC~\cite{CDFD0:2018qhk}.

\subsection{ATLAS measurements}

The higher Drell--Yan cross section at the LHC provides ATLAS with more events per unit integrated luminosity than at the Tevatron.  The larger quark-direction dilution in $pp$ collisions increases the PDF sensitivity of the mixing angle extraction.  The 2015 ATLAS 7 TeV measurement~\cite{ATLAS:2015ihy} used 4.8 fb$^{-1}$ of electron data and 4.6 fb$^{-1}$ of muon data.  Central--forward electron pairs were included alongside the central electron and muon channels, extending the dilepton rapidity coverage and reducing quark-direction dilution.  Despite having the fewest selected events, the forward-electron channel gave the smallest statistical uncertainty on the mixing angle among the three channels~\cite{ATLAS:2015ihy}.

The extraction used ATLAS-epWZ12 LO, a PDF fit prepared for this analysis using HERA inclusive DIS data and ATLAS 7 TeV W- and Z-boson measurements.  The combined result was $\sin^2\theta^\ell_{\rm eff}=0.23080\pm0.00120$, with a PDF uncertainty of $0.00090$ evaluated using this set~\cite{ATLAS:2015ihy}.

The 2018 preliminary ATLAS 8 TeV analysis~\cite{ATLAS:2018gqq} used 20.2 fb$^{-1}$ and retained the three channels, covering dilepton rapidities up to $|y|=3.6$.  The Born-level angular coefficients in the full lepton decay phase space were determined by fitting reconstructed $(\cos\theta,\phi)$ distributions in mass and rapidity bins with templates that included the detector response.  The mixing angle extraction used the predicted dependence of $A_4(m,|y|)$ on $\sin^2\theta^\ell_{\rm eff}$.  The predictions were obtained with \textsc{DYTURBO}, including electroweak form factors from \textsc{Dizet} 6.21 through \textsc{TauSpinner} event weights~\cite{ATLAS:2018gqq,RichterWas:2019TauSpinnerEW}.

The mixing angle fit used MMHT14 NNLO and profiled its eigenvectors, allowing the angular data to constrain the PDFs~\cite{ATLAS:2018gqq}.  The ATLAS 2018 8 TeV preliminary result is
\begin{equation}
\sin^2\theta^\ell_{\rm eff}=0.23140
\pm0.00021_{\rm stat}
\pm0.00024_{\rm PDF}
\pm0.00016_{\rm syst},
\end{equation}
with a total uncertainty of 0.00036.  The statistical and PDF contributions are comparable, and the PDF uncertainty is larger than those reported by CDF and D0.

\subsection{CMS measurements}

CMS also benefits from higher Drell--Yan event rates and faces similar limitations from quark-direction dilution and PDFs.  Unlike ATLAS, the 2018 CMS 8 TeV measurement~\cite{CMS:2018ktx} used only central electron and muon channels, with 19.6 fb$^{-1}$ of dielectron data and 18.8 fb$^{-1}$ of dimuon data.  The mixing angle was extracted from detector-level angular-weighted $A_{\rm FB}^{w}(m,|y|)$ distributions using \textsc{POWHEG} templates with NNPDF3.0 NLO replicas.  Giele--Keller reweighting used the measured mass and rapidity dependence to constrain these replicas, reducing the PDF uncertainty from $0.00057$ to $0.00031$, including the finite-ensemble contribution~\cite{CMS:2018ktx}.  The combined electron--muon result was
\begin{equation}
\sin^2\theta^\ell_{\rm eff}=0.23101
\pm0.00036_{\rm stat}
\pm0.00018_{\rm syst}
\pm0.00016_{\rm theo}
\pm0.00031_{\rm PDF},
\end{equation}
with a total uncertainty of 0.00053.  After reweighting, the statistical and PDF contributions were comparable.

The 2025  CMS 13 TeV analysis~\cite{CMS:2024ony} used 138 fb$^{-1}$ and added forward-electron channels, extending the dilepton rapidity coverage from $|y|<2.4$ to $|y|<3.4$.  These channels provided sensitivity in a region with smaller quark-direction dilution.  Electrons beyond the tracking acceptance were reconstructed using the forward calorimeters, including the forward hadron calorimeter beyond the crystal ECAL.  Their use required dedicated identification, efficiency measurements, energy-response corrections, and control of jet backgrounds~\cite{CMS:2024ony}.

The mixing angle was extracted by fitting detector-level angular-weighted $A_{\rm FB}^{w}$ in mass and rapidity bins using \textsc{POWHEG} MiNNLOPS simulation.  Weak corrections and mixing angle variations were evaluated with \textsc{Z\_ew-BMNNPV} using the $\sin^2\theta^\ell_{\rm eff}$ input scheme described in Section~\ref{sec:ew_input_schemes}.  The analysis used CT18Z NNLO Hessian PDFs, whose eigenvectors were profiled together with the mixing angle and the other nuisance parameters~\cite{CMS:2024ony}.

The baseline result was
\begin{equation}
\sin^2\theta^\ell_{\rm eff}=0.23152
\pm0.00010_{\rm stat}
\pm0.00015_{\rm exp}
\pm0.00008_{\rm theo}
\pm0.00027_{\rm PDF},
\end{equation}
with a total uncertainty of 0.00031 after accounting for correlations.  The statistical uncertainty was more than a factor of three smaller than at 8 TeV, leaving the PDF contribution as the dominant uncertainty.  Alternative PDF sets gave total uncertainties between 0.00024 and 0.00035~\cite{CMS:2024ony}.

CMS also published unfolded $A_4(m,|y|)$ distributions before QED final-state radiation, together with their covariance matrices.  Fits to these distributions and direct fits to the $(m,|y|,\cos\theta)$ distributions gave mixing angle results compatible with the angular-weighted $A_{\rm FB}^{w}$ extraction~\cite{CMS:2024ony}.

\subsection{LHCb measurements}

LHCb measured the dimuon asymmetry with both muons in $2.0<\eta_\mu<4.5$.  This forward acceptance reduces quark-direction dilution and gives good sensitivity to the mixing angle.  The 7 and 8 TeV measurements used 1 and 2 fb$^{-1}$, respectively, to determine $A_{\rm FB}(m_{\mu\mu})$~\cite{LHCb:2015jyu}.

The 2024 LHCb 13 TeV analysis~\cite{LHCb:2024ygc} used 5.4 fb$^{-1}$ recorded in 2016--2018 and measured $A_{\rm FB}$ in bins of the absolute muon pseudorapidity difference, $|\Delta\eta_{\mu\mu}|$, integrated over $66<m_{\mu\mu}<116$ GeV and the accepted dilepton rapidity range.  At negligible dilepton transverse momentum, $|\cos\theta_{\rm CS}|\simeq\tanh(|\Delta\eta_{\mu\mu}|/2)$.  This binning therefore separates events with different angular sensitivity to the mixing angle~\cite{LHCb:2024ygc}.

The measured asymmetries were corrected to the fiducial bare-muon level before comparison with theory.  The predictions used \textsc{Powheg-ewnlo} with NLO QCD and electroweak corrections in the $\sin^2\theta^\ell_{\rm eff}$ input scheme, supplemented by \textsc{Photos} and \textsc{Pythia} 8.  Baseline generation used NNPDF3.1 NLO.  The reported mixing angle was an arithmetic average of the extractions with NNPDF3.1, CT18, and MSHT20, all at NLO~\cite{LHCb:2024ygc}.

The result was
\begin{equation}
\sin^2\theta^\ell_{\rm eff}=0.23147
\pm0.00044_{\rm stat}
\pm0.00005_{\rm syst}
\pm0.00023_{\rm model}.
\end{equation}
The model uncertainty included a PDF contribution of 0.00021, smaller than the statistical uncertainty of 0.00044.  The precision was therefore mainly limited by statistics, and PDF profiling was not used.  At this level of  precision, there was less need to use the mass and dilepton rapidity dependence for additional PDF constraints~\cite{LHCb:2024ygc}.

\begin{table*}[!htbp]
\centering
\small
\renewcommand{\arraystretch}{0.99}
\setlength{\tabcolsep}{3pt}
\caption{Selected Drell--Yan measurements of $\sin^2\theta^\ell_{\rm eff}$, with their analysis methods and approximate kinematic coverage.  The publication year and integrated luminosity are shown for each measurement, using the larger luminosity when channels differ.  Uncertainties are given in units of $10^{-5}$.  Total uncertainties include contributions beyond the statistical and PDF components and account for the reported correlations.  The CDF and D0 entries refer to their combined electron and muon results.}
\label{tab:hadron_measurements}
\begin{tabular}{@{}c@{\hspace{10pt}}c@{}}
\toprule
\begin{minipage}[t]{0.19\textwidth}
\raggedright
\textbf{CDF 1.96 TeV~\cite{CDF:2016cei,CDFD0:2018qhk}}\par\vspace{1pt}
\textit{2016, $9.4\,\mathrm{fb}^{-1}$}\par\vspace{1pt}
\textit{Uncertainty}\par\vspace{1pt}
\begin{tabular}{@{}lr@{}}
Stat. & 43 \\
PDF & 16 \\
Total & 46 \\
\end{tabular}
\end{minipage} &
\begin{minipage}[t]{\dimexpr0.81\textwidth-10pt\relax}
\begin{tabularx}{\linewidth}[t]{@{}>{\raggedright\arraybackslash}p{0.36\linewidth}>{\raggedright\arraybackslash}p{0.32\linewidth}>{\raggedright\arraybackslash}X@{}}
\textit{Observable} & \textit{EW treatment} & \textit{PDF treatment} \\
Angular-weighted $A_{\rm FB}(m)$\newline (unfolded) & \textsc{ZFITTER}/\textsc{Dizet} EW corrections & NNPDF3.0 NNLO replicas\newline GK reweighting \\
\cmidrule(lr){1-3}
\textit{Channel} & \textit{Lepton acceptance} & \textit{Dilepton rapidity} \\
$ee$ (central--central, central--forward) & Central $0.05<|\eta|<1.05$\newline Forward $1.2<|\eta|<2.8$ & $|y|<1.7$ \\
$\mu\mu$ (central--central) & Trigger muon $|\eta|<1$\newline Partner acceptance to $|\eta|\simeq1.5$ & $|y|<1$
\end{tabularx}
\end{minipage} \\
\midrule
\begin{minipage}[t]{0.19\textwidth}
\raggedright
\textbf{D0 1.96 TeV~\cite{D0:2017ekd,CDFD0:2018qhk}}\par\vspace{1pt}
\textit{2018, $9.7\,\mathrm{fb}^{-1}$}\par\vspace{1pt}
\textit{Uncertainty}\par\vspace{1pt}
\begin{tabular}{@{}lr@{}}
Stat. & 35 \\
PDF & 19 \\
Total & 40 \\
\end{tabular}
\end{minipage} &
\begin{minipage}[t]{\dimexpr0.81\textwidth-10pt\relax}
\begin{tabularx}{\linewidth}[t]{@{}>{\raggedright\arraybackslash}p{0.36\linewidth}>{\raggedright\arraybackslash}p{0.32\linewidth}>{\raggedright\arraybackslash}X@{}}
\textit{Observable} & \textit{EW treatment} & \textit{PDF treatment} \\
$A_{\rm FB}(m)$\newline (detector level) & \textsc{ZFITTER}/\textsc{Dizet} EW corrections & NNPDF3.0 NNLO reference\newline No reweighting \\
\cmidrule(lr){1-3}
\textit{Channel} & \textit{Lepton acceptance} & \textit{Dilepton rapidity} \\
$ee$ (central--central, central--forward, forward--forward) & Central $|\eta|<1.1$\newline Forward $1.5<|\eta|<3.2$ & No separate cut \\
$\mu\mu$ (central--central) & Both $|\eta|<1.8$\newline At least one $|\eta|<1.6$ & No separate cut
\end{tabularx}
\end{minipage} \\
\midrule
\begin{minipage}[t]{0.19\textwidth}
\raggedright
\textbf{ATLAS 8 TeV~\cite{ATLAS:2018gqq}}\par\textit{preliminary}\par\vspace{1pt}
\textit{2018, $20.2\,\mathrm{fb}^{-1}$}\par\vspace{1pt}
\textit{Uncertainty}\par\vspace{1pt}
\begin{tabular}{@{}lr@{}}
Stat. & 21 \\
PDF & 24 \\
Total & 36 \\
\end{tabular}
\end{minipage} &
\begin{minipage}[t]{\dimexpr0.81\textwidth-10pt\relax}
\begin{tabularx}{\linewidth}[t]{@{}>{\raggedright\arraybackslash}p{0.36\linewidth}>{\raggedright\arraybackslash}p{0.32\linewidth}>{\raggedright\arraybackslash}X@{}}
\textit{Observable} & \textit{EW treatment} & \textit{PDF treatment} \\
$A_4(m,|y|)$\newline (unfolded) & \textsc{ZFITTER}/\textsc{Dizet} EW corrections & MMHT14 NNLO Hessian\newline Profiling \\
\cmidrule(lr){1-3}
\textit{Channel} & \textit{Lepton acceptance} & \textit{Dilepton rapidity} \\
$ee$, $\mu\mu$ (central--central) & $|\eta_\ell|<2.4$ & $|y|<2.5$ \\
$ee$ (central--forward) & Central $|\eta_e|<2.4$\newline Forward $2.5<|\eta_e|<4.9$ & $1.6<|y|<3.6$
\end{tabularx}
\end{minipage} \\
\midrule
\begin{minipage}[t]{0.19\textwidth}
\raggedright
\textbf{CMS 8 TeV~\cite{CMS:2018ktx}}\par\vspace{1pt}
\textit{2018, $19.6\,\mathrm{fb}^{-1}$}\par\vspace{1pt}
\textit{Uncertainty}\par\vspace{1pt}
\begin{tabular}{@{}lr@{}}
Stat. & 36 \\
PDF & 31 \\
Total & 53 \\
\end{tabular}
\end{minipage} &
\begin{minipage}[t]{\dimexpr0.81\textwidth-10pt\relax}
\begin{tabularx}{\linewidth}[t]{@{}>{\raggedright\arraybackslash}p{0.36\linewidth}>{\raggedright\arraybackslash}p{0.32\linewidth}>{\raggedright\arraybackslash}X@{}}
\textit{Observable} & \textit{EW treatment} & \textit{PDF treatment} \\
Angular-weighted $A_{\rm FB}(m,|y|)$\newline (detector level) & Effective Born approximation
& NNPDF3.0 NLO replicas\newline GK reweighting \\
\cmidrule(lr){1-3}
\textit{Channel} & \textit{Lepton acceptance} & \textit{Dilepton rapidity} \\
$ee$, $\mu\mu$ (central--central) & $|\eta_\ell|<2.4$ & $|y|<2.4$ \\
\end{tabularx}
\end{minipage} \\
\midrule
\begin{minipage}[t]{0.19\textwidth}
\raggedright
\textbf{CMS 13 TeV~\cite{CMS:2024ony}}\par\vspace{1pt}
\textit{2025, $138\,\mathrm{fb}^{-1}$}\par\vspace{1pt}
\textit{Uncertainty}\par\vspace{1pt}
\begin{tabular}{@{}lr@{}}
Stat. & 10 \\
PDF & 27 \\
Total & 31 \\
\end{tabular}
\end{minipage} &
\begin{minipage}[t]{\dimexpr0.81\textwidth-10pt\relax}
\begin{tabularx}{\linewidth}[t]{@{}>{\raggedright\arraybackslash}p{0.36\linewidth}>{\raggedright\arraybackslash}p{0.32\linewidth}>{\raggedright\arraybackslash}X@{}}
\textit{Observable} & \textit{EW treatment} & \textit{PDF treatment} \\
Angular-weighted $A_{\rm FB}(m,|y|)$\textsuperscript{a}\newline (detector level) & $\sin^2\theta^\ell_{\rm eff}$ input scheme & CT18Z NNLO Hessian\newline Profiling \\
\cmidrule(lr){1-3}
\textit{Channel} & \textit{Lepton acceptance} & \textit{Dilepton rapidity} \\
$ee$, $\mu\mu$ (central--central) & $|\eta_e|<2.5$, $|\eta_\mu|<2.4$ & $|y|<2.4$ \\
$ee$ (central--forward, ECAL) & Central $|\eta_e|<2.5$\newline Forward $2.5<|\eta_e|<2.87$ & $1.2<|y|<2.7$ \\
$ee$ (central--forward, HCAL) & Central $1.57<|\eta_e|<2.5$\newline Forward $3.14<|\eta_e|<4.36$ & $2.0<|y|<3.4$
\end{tabularx}
\end{minipage} \\
\midrule
\begin{minipage}[t]{0.19\textwidth}
\raggedright
\textbf{LHCb 13 TeV~\cite{LHCb:2024ygc}}\par\vspace{1pt}
\textit{2024, $5.4\,\mathrm{fb}^{-1}$}\par\vspace{1pt}
\textit{Uncertainty}\par\vspace{1pt}
\begin{tabular}{@{}lr@{}}
Stat. & 44 \\
PDF & 21 \\
Total & 50 \\
\end{tabular}
\end{minipage} &
\begin{minipage}[t]{\dimexpr0.81\textwidth-10pt\relax}
\begin{tabularx}{\linewidth}[t]{@{}>{\raggedright\arraybackslash}p{0.36\linewidth}>{\raggedright\arraybackslash}p{0.32\linewidth}>{\raggedright\arraybackslash}X@{}}
\textit{Observable} & \textit{EW treatment} & \textit{PDF treatment} \\
$A_{\rm FB}(|\Delta\eta_{\mu\mu}|)$\newline (unfolded) & $\sin^2\theta^\ell_{\rm eff}$ input scheme & Average of CT18, MSHT20, NNPDF3.1 NLO\newline No profiling \\
\cmidrule(lr){1-3}
\textit{Channel} & \textit{Lepton acceptance} & \textit{Dilepton rapidity} \\
$\mu\mu$ (forward--forward) & $2<\eta_\mu<4.5$ & No separate cut
\end{tabularx}
\end{minipage} \\
\bottomrule
\end{tabular}
\par\vspace{3pt}
\begin{minipage}{\textwidth}
\footnotesize
\textsuperscript{a} CMS also publishes unfolded pre-FSR $A_4(m,|y|)$ distributions.  The tabulated uncertainties refer to the detector-level angular-weighted $A_{\rm FB}$ result.
\end{minipage}
\end{table*}

The uncertainty contributions in Table~\ref{tab:hadron_measurements} show the different limitations of these measurements.  The Tevatron combination and the LHCb 13 TeV measurement remain mainly limited by statistics.  The ATLAS 8 TeV result has comparable statistical and PDF uncertainties, whereas PDFs dominate the CMS 13 TeV uncertainty.
To provide an analytic understanding of this limitation,
Section~\ref{sec:pdf_dependence} examines which PDF combinations produce differences between the predicted Drell--Yan asymmetries.  Section~\ref{sec:common_lhc_interpretation} discusses a common interpretation of the LHC measurements and estimates their PDF correlations and the expected precision of a combination.

\section{PDF flavor structure and Drell--Yan asymmetries}
\label{sec:pdf_dependence}

As larger Drell--Yan datasets at the LHC reduce the statistical uncertainties, understanding PDF uncertainties becomes increasingly important for more precise weak-mixing-angle determinations.  The CMS 13~TeV Run~2 analysis already illustrates this situation.  With 138~fb$^{-1}$ of data, the PDF uncertainty is $27\times10^{-5}$, compared with a total uncertainty of $31\times10^{-5}$~\cite{CMS:2024ony}.  In this section, we therefore examine the flavor structure of modern PDF sets using a simplified leading-order Z-pole calculation.  By distinguishing the contributions of quark-flavor production weights and quark-direction dilution, we connect the PDF-set dependence of the predicted asymmetry to the underlying PDF combinations and identify momentum-fraction regions where additional constraints could improve future measurements.

At leading order in $pp$ collisions, let $x_{\rm hi,lo}=(m/\sqrt{s})e^{\pm|y|}$ for a dilepton pair with mass $m$ and rapidity $y$, and write $q_i=f_q(x_i,\mu_F^2)$ and $\bar q_i=f_{\bar q}(x_i,\mu_F^2)$ at factorization scale $\mu_F\simeq m$.  When the dilepton boost is used to estimate the quark direction, the combination $q_{\rm hi}\bar q_{\rm lo}$ corresponds to a correct assignment, whereas $\bar q_{\rm hi}q_{\rm lo}$ corresponds to a reversed assignment.  Both combinations contribute to the total rate, but the reversed assignment changes the sign of the forward--backward difference and therefore dilutes the asymmetry~\cite{Ball:2022DYAFB}.  Accounting for both contributions, the pure-Z, leading-order asymmetry near the Z pole is
\begin{equation}
A_{\rm FB}^{\rm LO}(P,s_0^2)=
\frac{\displaystyle\sum_q w_q\widehat A_{\rm FB}^{q} L_{A,q}(P)}
     {\displaystyle\sum_q w_q L_{S,q}(P)}.
\label{eq:lo_pdf_afb}
\end{equation}
Here $P$ denotes a PDF set and the sum runs over $q=u,d,s,c,b$.  The luminosities are defined over the chosen $|y|$ interval as
\[
L_{S,q}=\int d|y|\,\bigl[q_{\rm hi}\bar q_{\rm lo}+\bar q_{\rm hi}q_{\rm lo}\bigr],\qquad
L_{A,q}=\int d|y|\,\bigl[q_{\rm hi}\bar q_{\rm lo}-\bar q_{\rm hi}q_{\rm lo}\bigr].
\]
The flavor weight is $w_q=v_q^2+a_q^2$, and the undiluted partonic asymmetry is $\widehat A_{\rm FB}^{q}=(3/4)A_\ell A_q$, with $A_f=2v_fa_f/(v_f^2+a_f^2)$.  We use the tree-level couplings $v_f=T_3^f-2Q_fs_0^2$ and $a_f=T_3^f$, where $T_3^f$ and $Q_f$ are the third component of weak isospin and the electric charge in units of the positron charge, and $s_0^2$ is the common reference value of $s_{\rm eff}^2$. The discussion here focuses on $pp$ collisions, but the same analysis can be carried out for $p\bar p$ collisions with appropriately defined $L_{S,q}$ and $L_{A,q}$.

\begin{figure*}[!htb]
\centering
\includegraphics[width=5.8in]{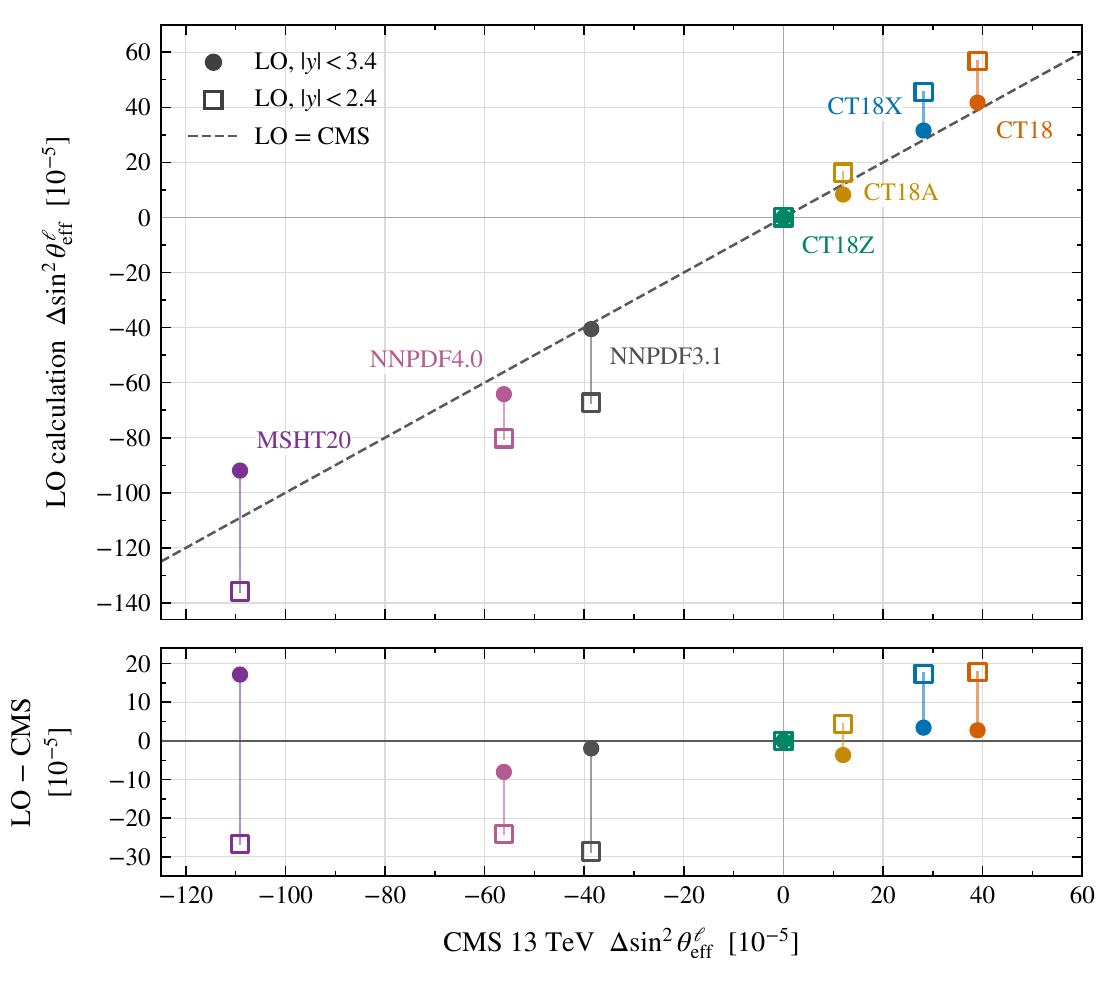}
\caption{Comparison of PDF-set differences in $\sin^2\theta^\ell_{\rm eff}$ estimated from LO Z-pole asymmetries (vertical axis) with those from the unprofiled CMS 13~TeV $A_{\rm FB}^{w}$ fit (horizontal axis).  The CMS values are digitized from Fig.~11 of Ref.~\cite{CMS:2024ony}.  All differences are defined relative to CT18Z and given in units of $10^{-5}$.  Filled circles and open squares correspond to LO integrations over $|y|<3.4$ and $|y|<2.4$, respectively.  The connecting segments show the change between these rapidity ranges.  The dashed line denotes equality between the LO and CMS results, and the lower panel shows LO minus CMS.  The LO calculation uses $m=\mu_F=m_Z$, $\sqrt{s}=13$~TeV, and $s_0^2=0.23152$, with the CT18Z slope evaluated separately for each rapidity range.}
\label{Fig_cms_lo_pdf_trend}
\end{figure*}

Despite its simplifying assumptions, the LO Z-pole calculation can help explain PDF-set differences in experimental extractions of $\sin^2\theta^\ell_{\rm eff}$.  To compare with the {\it unprofiled} CMS 13~TeV $A_{\rm FB}^{w}$-fit results~\cite{CMS:2024ony}, we express the LO asymmetry difference from CT18Z as a mixing-angle shift at fixed measured asymmetry, $\Delta s_{\rm eff}^2\simeq-\Delta A_{\rm FB}^{\rm LO}/(\partial A_{\rm FB}^{\rm LO}/\partial s_{\rm eff}^2)$.  The derivative is evaluated with CT18Z at the common input mixing angle $s_0^2$.  Figure~\ref{Fig_cms_lo_pdf_trend} shows reasonable agreement with the CMS shifts for $|y|<3.4$.  This calculation omits experimental acceptance and fit-bin weighting.  Results for $|y|<2.4$ illustrate rapidity-range dependence with unchanged PDF-set ordering.

\begin{figure*}[!htb]
\centering
\includegraphics[width=\textwidth]{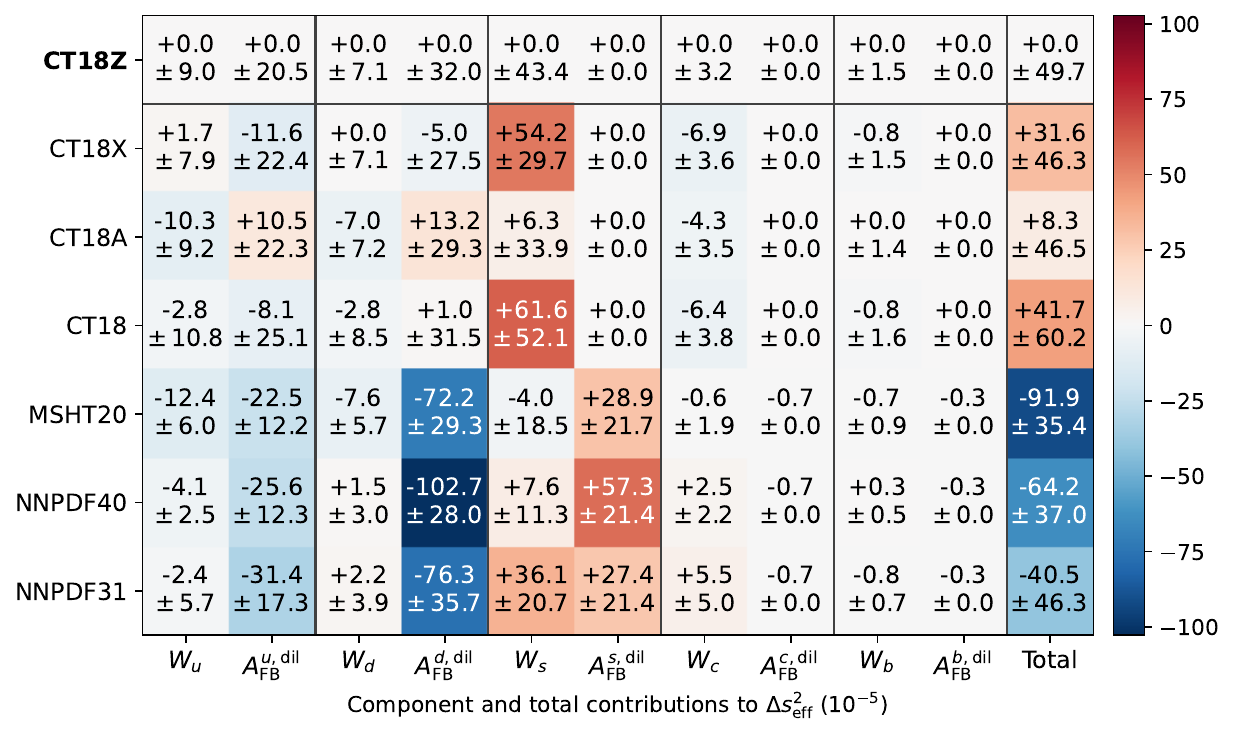}
\caption{Flavor contributions to mixing-angle shifts $\Delta s_{\rm eff}^2$ relative to CT18Z, in units of $10^{-5}$. Contributions from production-weight ($W_q$) and diluted-asymmetry ($A_{\rm FB}^{q,\rm dil}$) changes are shown separately. The asymmetry shifts $\Delta A_{\rm FB}$ are converted to mixing-angle shifts $\Delta s_{\rm eff}^2$ using the CT18Z slope $\partial A_{\rm FB}^{\rm LO}/\partial s_{\rm eff}^2\simeq-1.35$ for $|y|<3.4$, as in Fig.~\ref{Fig_cms_lo_pdf_trend}. Cells show central values and symmetric 68\% CL Hessian PDF uncertainties.  The Total column gives their sum with its correlated uncertainty.  CT18-family strange-asymmetry contributions vanish because $s=\bar s$ is imposed.}
\label{Fig_cms_lo_flavor_components}
\end{figure*}

The approximate agreement in Fig.~\ref{Fig_cms_lo_pdf_trend} motivates using the LO calculation to examine the origin of the PDF-set differences.  We rewrite Eq.~\ref{eq:lo_pdf_afb} as a weighted mean of the diluted asymmetries for the individual quark flavors,
\begin{equation}
A_{\rm FB}^{\rm LO}=\frac{\sum_q W_q A_{\rm FB}^{q,\rm dil}}{\sum_q W_q},
\label{eq:lo_pdf_weighted_mean}
\end{equation}
where $W_q=w_qL_{S,q}$ is the production weight and $A_{\rm FB}^{q,\rm dil}=\widehat A_{\rm FB}^{q}L_{A,q}/L_{S,q}$ is the diluted asymmetry for flavor $q$.  Taking CT18Z as the reference, we evaluate the contribution of each component to the mixing-angle shift $\Delta s_{\rm eff}^2$ for a given comparison PDF set.  For each component, we calculate the shift caused by replacing its CT18Z value with the corresponding value from that set.  We average this shift over all possible orders of replacing the ten components (the Shapley value)~\cite{Shapley:1953value}.  Figure~\ref{Fig_cms_lo_flavor_components} shows the contributions in mixing-angle units.

The flavor decomposition in Fig.~\ref{Fig_cms_lo_flavor_components} shows two distinct patterns.  Within the CT18 family, the positive shifts of CT18 and CT18X relative to CT18Z result primarily from the strange-quark production weight.  For CT18, this contribution is $+61.6\times10^{-5}$, partly offset by other contributions to give a total shift of $+41.7\times10^{-5}$.  In these input CT sets, $s=\bar s$ and hence $L_{A,s}=0$.  Strange production adds to the denominator without contributing to the forward--backward numerator.  A smaller strange luminosity therefore reduces this dilution and raises the total asymmetry.

This strangeness pattern is consistent with the treatment of the precision ATLAS \emph{7~TeV} $W/Z$ data in the CT18 analysis~\cite{Hou:2019efy}.  These data are included in CT18A and CT18Z, but not in CT18 or CT18X.  They favor enhanced strangeness at the electroweak scale, as reflected in the larger strange luminosities of CT18A/Z in the LO calculation.

For the MSHT20 and NNPDF sets shown, the largest shift relative to CT18Z comes from the diluted $d$-quark asymmetry and is negative.  The diluted $u$-quark asymmetry gives a smaller shift in the same direction.  Their nonzero strange-quark asymmetries give positive contributions that partly compensate the reduced light-quark asymmetry.  Production-weight changes further modify the shifts, but the net LO asymmetries remain below CT18Z.

Even the LO calculation of $A_{\rm FB}$ is rather involved for a quick comparison of PDF sets, as it requires PDF products, flavor-dependent weights, and rapidity integration.  Simple combinations of PDFs would allow the main differences to be examined more directly.  To show how the quark--antiquark imbalance enters the asymmetry, we introduce the sum $S_q=q+\bar q$ and difference $V_q=q-\bar q$.  The luminosities can then be written as
\begin{equation}
\begin{aligned}
L_{A,q}&=\frac{1}{2}\int d|y|\,\bigl[V_{q,\rm hi}S_{q,\rm lo}-S_{q,\rm hi}V_{q,\rm lo}\bigr],\\
L_{S,q}&=\frac{1}{2}\int d|y|\,\bigl[S_{q,\rm hi}S_{q,\rm lo}-V_{q,\rm hi}V_{q,\rm lo}\bigr].
\end{aligned}
\label{eq:pdf_valence_luminosities}
\end{equation}

The strange density $s+\bar s$ provides a simple way to examine the production-weight differences within the CT18 family.  These sets impose $s=\bar s$, so $V_s=0$.  Equation~\ref{eq:pdf_valence_luminosities} then gives $L_{A,s}=0$, while $L_{S,s}$ depends only on the product of $s+\bar s$ on the two incoming legs.  Strange production therefore affects only the denominator and dilutes the asymmetry.  Comparing $s+\bar s$ over the relevant $x$ range helps identify the PDF differences responsible for this effect.

The summed quark--antiquark difference $Q-\bar Q$, where $Q=u+d+s+c+b$, provides a simple quantity for examining the PDF differences behind the diluted quark asymmetries.  To connect it to $A_{\rm FB}$, we first neglect the quark--antiquark difference on the low-$x$ leg, $V_{q,\rm lo}\simeq0$.  The differential asymmetry then becomes
\begin{equation}
A_{\rm FB}^{\rm LO}(y)\simeq
\frac{\sum_q w_q\widehat A_{\rm FB}^{q} S_{q,\rm lo}V_{q,\rm hi}}
     {\sum_q w_q S_{q,\rm lo}S_{q,\rm hi}}.
\label{eq:lo_pdf_valence_limit}
\end{equation}
Further ignoring flavor differences in the low-$x$ sea and electroweak weights gives the very rough relation $A_{\rm FB}^{\rm LO}(y)\propto(Q-\bar Q)_{\rm hi}/(Q+\bar Q)_{\rm hi}$.  This simplification motivates using $Q-\bar Q$ as a simple diagnostic of the high-$x$ PDF differences entering the asymmetry numerator.  The low-$x$ approximation in Eq.~\ref{eq:lo_pdf_valence_limit} improves at forward rapidity and fails at $y=0$, where the two terms in $L_{A,q}$ cancel exactly.  For CT18Z, setting $V_{q,\rm lo}=0$ in both luminosities overestimates the integrated LO asymmetry by 7.8\% for $|y|<3.4$, and by 1.0\% for $2<|y|<3.4$.

\begin{figure*}[!htb]
\centering
\includegraphics[width=\textwidth]{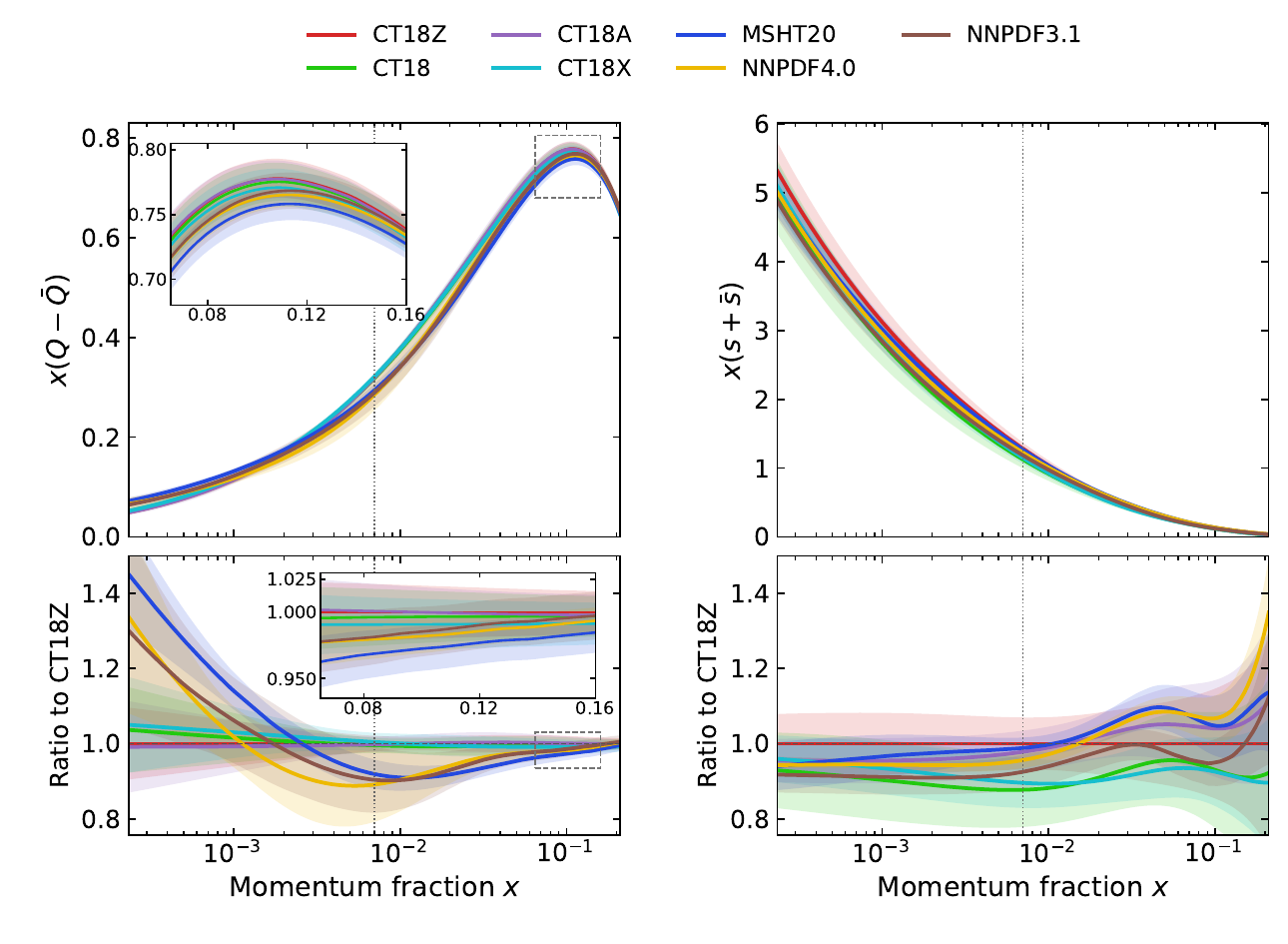}
\caption{PDF combinations used to examine the quark--antiquark imbalance and strange density entering the LO asymmetry.  The upper panels show $x(Q-\bar Q)$ (left), where $Q=u+d+s+c+b$, and $x(s+\bar s)$ (right) at $\mu_F=m_Z$.  Lower panels show ratios to the CT18Z prediction.  Curves correspond to the NNLO PDF sets in the legend, with symmetric 68\% CL uncertainty bands.  The left-hand insets enlarge the peak region near $x=0.1$ and the corresponding ratios.  The plotted range covers both momentum-fraction branches for LO Z-pole production at $\sqrt{s}=13$~TeV and $|y|<3.4$, separated by the dotted line at $x=m_Z/\sqrt{s}$.}
\label{Fig_pdf_valence}
\end{figure*}
\FloatBarrier

Figure~\ref{Fig_pdf_valence} compares $Q-\bar Q$ and $s+\bar s$ over the momentum-fraction range sampled by the LO calculation.  In the left panel, the CT18-family sets have similar $Q-\bar Q$ distributions on the high-$x$ branch, whereas MSHT20 and the NNPDF sets give smaller values over much of $x\sim10^{-2}$--$10^{-1}$.  This pattern is consistent with the reduced light-quark asymmetry found in the flavor decomposition.  The ordering reverses at the lowest $x$, but the quark--antiquark imbalance relative to the total density is much smaller there than on the high-$x$ leg.  This suppresses the low-$x$ valence contribution to the asymmetry.

In the right panel, CT18 and CT18X have smaller $s+\bar s$ than CT18Z on both the low- and high-$x$ branches.  Together with their similar high-$x$ $Q-\bar Q$ distributions, this supports the interpretation that the positive CT18 and CT18X shifts are mainly due to the strange production weight.

These results motivate better constraints on the strange density over both momentum-fraction branches and on the high-$x$ quark--antiquark differences at $x\sim10^{-2}$--$10^{-1}$.  The measured mass and rapidity dependence of $A_{\rm FB}$ or $A_4$ can constrain PDFs through Hessian profiling or replica reweighting.  These constraints reduce both the PDF uncertainty and the spread of extracted mixing angles, although some differences between PDF sets remain~\cite{CMS:2024ony}.

W charge asymmetries and W/Z cross-section ratios can provide complementary PDF constraints.  An LHC simulation study showed that adding the charged-lepton asymmetry in W decays to an $A_{\rm FB}$ analysis reduces the PDF uncertainty in $\sin^2\theta^\ell_{\rm eff}$~\cite{Bodek:2015ljm}. 
In a recent, independent phenomenological reinterpretation~\cite{Bodek:2025kxl}, the published CMS 13~TeV $A_4$ measurement was fitted together with the CMS W-decay lepton charge asymmetry~\cite{CMS:2020cph} and W/Z cross-section ratios~\cite{CMS:2024myi}. This separate study demonstrated that the additional constraints successfully reduced 
both the PDF uncertainty and the spread of extracted mixing angles among PDF sets.
For CT18Z, this study obtained $\sin^2\theta^\ell_{\rm eff}=0.23154\pm0.00024$, compared with the published CMS result of $0.23152\pm0.00031$~\cite{Bodek:2025kxl,CMS:2024ony}.

\FloatBarrier

\section{Toward a common reinterpretation of LHC measurements}
\label{sec:common_lhc_interpretation}

ATLAS, CMS, and LHCb measure the same quantity, $\sin^2\theta^\ell_{\rm eff}$, but extract it using different electroweak calculations, PDF sets, and PDF constraints.  A difference between the published values can therefore include a shift from these inputs.  Testing statistical compatibility requires these shifts and the correlated PDF uncertainties to be evaluated.  A common reinterpretation of the angular measurements would provide the basis for this comparison and for a combination.  In this section, we discuss the requirements for such a reinterpretation and use a simple estimate to illustrate how PDF correlations limit the gain in precision from combining the current measurements.

The Tevatron combination provides an example of a common reinterpretation.  The input central values were expressed using common NNPDF3.0 NNLO PDFs and a \textsc{ZFITTER}-based electroweak prescription.  Dedicated template comparisons were used to evaluate the required mixing angle shifts and the uncertainties of the corrections.  The corrected mixing angle results were then combined with BLUE, assuming fully correlated PDF uncertainties.  The combination was performed at the level of the extracted mixing angles, without a joint fit to the CDF and D0 angular distributions~\cite{CDFD0:2018qhk}.

Combining only the extracted mixing angles and their covariance does not automatically use all the PDF information in the measured angular distributions~\cite{Dado:2026blue}.  The CDF data, for example, can constrain PDF parameters that also affect the D0 mixing angle and its uncertainty.  Changing the PDF set or electroweak calculation also requires the central values and uncertainties to be reevaluated.  These refits require unfolded asymmetry measurements and their covariance matrices.  D0 did not provide tabulated unfolded $A_{\rm FB}$ values or their covariance~\cite{D0:2017ekd,CDFD0:2018qhk}, while CDF provided fully corrected $A_{\rm FB}(m)$ values with bin-by-bin errors but not the full covariance used in the fit~\cite{CDF:2016cei}.

A more complete approach is to publish unfolded $A_{\rm FB}$ or $A_4$ measurements with covariance matrices and use them in a joint fit.  The fit uses a common electroweak calculation and PDF set, with shared PDF and theory nuisance parameters.  Constraints from each measurement are shared across experiments, and the PDF correlations follow from their responses to the same nuisance parameters.  The fit can test the consistency of the data and determine a common $\sin^2\theta^\ell_{\rm eff}$.  The same measurements can be refitted when the PDFs or electroweak calculations are updated.

Early LHC measurements did not publish unfolded angular distributions with full covariance matrices, making a common reinterpretation difficult.  CMS and LHCb provide these inputs for their 13~TeV measurements~\cite{CMS:2024ony,LHCb:2024ygc}.  For ATLAS, the preliminary 8~TeV mixing angle result does not provide the distributions and covariance needed for a refit~\cite{ATLAS:2018gqq}.  However, the published unfolded triple-differential Drell--Yan cross sections and their correlated uncertainties can be used to extract the mixing angle~\cite{ATLAS:2017dy3d}.  A joint analysis of the CMS 13~TeV, LHCb 13~TeV, and ATLAS 8~TeV datasets is therefore a useful first step toward a common LHC reinterpretation and combination.  The omitted earlier measurements are less precise and are expected to have little effect on the combined precision.

A joint fit can retain the observable and lepton definition of each published measurement.  For example, the CMS $A_4(m,|y|)$ values refer to leptons before QED FSR, while the LHCb $A_{\rm FB}(|\Delta\eta_{\mu\mu}|)$ values use bare muons, without recombining radiated photons, within the fiducial acceptance~\cite{CMS:2024ony,LHCb:2024ygc}.  Predictions are evaluated in the corresponding bins and phase space, with photon radiation treated according to each definition.  The fit uses these predictions and the experimental covariance matrices to determine a common mixing angle and shared PDF nuisance parameters.

To illustrate the possible gain in precision from combining the current LHC measurements, we make a simple estimate of the expected combined uncertainty.  The PDF correlations are evaluated using the simplified Z-pole calculation of Section~\ref{sec:pdf_dependence}, including the main lepton transverse-momentum and pseudorapidity cuts for CMS and LHCb at 13~TeV and ATLAS at 8~TeV~\cite{CMS:2024ony,ATLAS:2018gqq,LHCb:2024ygc}.  The acceptance integration assumes massless leptons and zero Z-boson transverse momentum.  For each PDF set, the three inclusive fiducial asymmetries are evaluated with the same Hessian error members, and their variations are converted to mixing angle shifts using the corresponding central slopes.  The five unprofiled NNLO PDF sets CT18, CT18Z, NNPDF3.1, NNPDF4.0, and MSHT20 give correlations in the range of
\begin{equation}
\begin{aligned}
\rho_{\rm CMS,ATLAS}&\simeq 0.835\text{--}0.962,\\
\rho_{\rm CMS,LHCb}&\simeq 0.506\text{--}0.642,\\
\rho_{\rm ATLAS,LHCb}&\simeq 0.717\text{--}0.768.
\end{aligned}
\label{eq:lhc_pdf_correlation}
\end{equation}
The ranges show the spread across the five PDF sets.  The stronger CMS--ATLAS correlation reflects their more similar kinematic coverage, while the forward LHCb acceptance gives a more distinct PDF response.  We use these LO correlations with the published PDF and total uncertainties and treat the remaining uncertainty as uncorrelated between experiments.  A simple estimate using the best linear unbiased estimator (BLUE)~\cite{Lyons:1988rp} then gives an expected combined uncertainty of $(26.8\text{--}27.5)\times10^{-5}$, compared with $31\times10^{-5}$ for CMS alone~\cite{CMS:2024ony}. 
We emphasize that this BLUE calculation is an illustrative projection to demonstrate the impact of PDF correlations, rather than an official experimental combination.
The improvement of 11--14\% illustrates the limited gain when the dominant PDF uncertainties are strongly correlated.

The W charge asymmetries and W/Z ratios discussed in Section~\ref{sec:pdf_dependence} could provide additional PDF constraints and improve the precision beyond this estimate.  These measurements can be included together with the unfolded angular distributions in a joint fit.  The fit should account for correlations among these data and any overlap with measurements already included in the input PDF fit.  The resulting PDF constraints can then be shared consistently across experiments.

\section{Outlook}

The full LHC Run 3 dataset at 13.6 TeV and the future HL-LHC dataset will provide larger Drell--Yan samples than those used in the current measurements.  These samples will reduce the statistical uncertainty in $\sin^2\theta^\ell_{\rm eff}$.  However, PDF uncertainty already dominates the most precise hadron-collider determination.  If this uncertainty remains at its present level, additional statistics will give only a limited improvement in the total precision.

Preserving unfolded $A_{\rm FB}$ or $A_4$ distributions with their experimental covariance matrices and observable definitions is therefore important for future mixing angle extractions.  These distributions can be refitted with updated PDFs or electroweak predictions.  They can also be used in joint fits across experiments and with the W charge asymmetries~\cite{CMS:2020cph} or W/Z ratios~\cite{CMS:2024myi} discussed in Section~\ref{sec:pdf_dependence}.

Larger data samples are particularly useful for measurements with smaller event yields.  One example is a determination of $\sin^2\theta^\ell_{\rm eff}$ from the lepton forward--backward asymmetry in $Z+b$ production.  The relevant Compton subprocesses are $gb\rightarrow Zb$ and $g\bar b\rightarrow Z\bar b$, followed by $Z\rightarrow\ell^+\ell^-$~\cite{delAguila:2002pr,delAguila:2005afb}.  Their contributions to the Collins--Soper asymmetry largely cancel when the bottom and antibottom PDFs are nearly equal.  A charge-sensitive asymmetry can instead be defined using the recoiling jet in the dilepton rest frame.  The forward axis is chosen opposite to the jet for b events and along the jet for $\bar b$ events, allowing their asymmetries to be combined~\cite{delAguila:2002pr}.

Using the jet direction avoids estimating the quark direction from the longitudinal boost, while bottom tagging selects a specific quark flavor.  These features can reduce the PDF sensitivity associated with direction dilution and the light-flavor mixture, providing a complementary determination of $\sin^2\theta^\ell_{\rm eff}$.  Residual dependence on the bottom and gluon PDFs and on additional production channels remains.  The smaller event yield, together with losses from bottom tagging and charge identification, makes large data samples essential.  Charge misidentification dilutes the asymmetry, so reliable bottom-charge determination and calibration of its efficiency and misidentification rate are also required~\cite{delAguila:2005afb}.

Larger data samples could also enable a test of the running of the weak mixing angle above the Z pole through high-mass Drell--Yan production~\cite{Amoroso:2023uux}.  The relevant quantity is the $\overline{\rm MS}$ parameter
\begin{equation}
\hat s_W^2(\mu)\equiv\sin^2\theta_W^{\overline{\rm MS}}(\mu)
=\frac{\hat g^{\prime\,2}(\mu)}
       {\hat g^2(\mu)+\hat g^{\prime\,2}(\mu)},
\label{eq:msbar_weak_angle}
\end{equation}
where $\hat g$ and $\hat g^\prime$ are the renormalized $SU(2)_L$ and $U(1)_Y$ couplings at scale $\mu$.  This parameter differs from $\sin^2\theta^\ell_{\rm eff}$ measured near the Z pole.  At the Z scale, the two are related by a finite electroweak form factor, $\sin^2\theta^\ell_{\rm eff}=\mathrm{Re}[\hat\kappa_\ell(m_Z)]\hat s_W^2(m_Z)$~\cite{PDG2026EW}.

\begin{figure*}
\includegraphics[width=6.4in,height=3.1in]{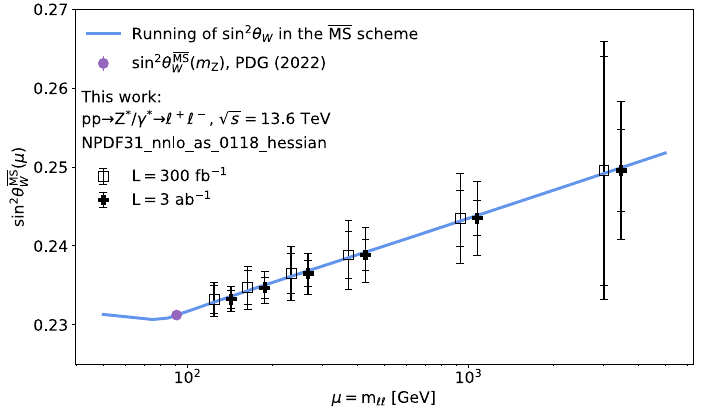}
\caption{Projected tests of the running $\overline{\rm MS}$ parameter $\hat s_W^2(\mu)$ in high-mass Drell--Yan production, from Ref.~\cite{Amoroso:2023uux}.  The blue curve shows SM running, and the violet point gives the Z-scale experimental reference expressed in the same scheme.  The study assumes 13.6 TeV collisions with 300 fb$^{-1}$ (Run 3) and 3000 fb$^{-1}$ (HL-LHC), and SM running of $\alpha_{\rm EM}^{\overline{\rm MS}}$.  The two scenarios are displaced horizontally for clarity.  Outer errors are total projected uncertainties.  Inner errors include statistical and experimental contributions only.}
\label{Fig_10}
\vspace{-0.1in}
\end{figure*}

The running parameter can be extracted at scales $\mu\sim m$ using predictions calculated in the same renormalization scheme.  Figure~\ref{Fig_10} shows projected sensitivities for Run 3 and the HL-LHC from a fit to cross sections binned in dilepton mass, rapidity, and forward or backward lepton direction~\cite{Amoroso:2023uux}.  The fit uses both rate and asymmetry information to determine a mixing angle in each mass interval.  These measurements complement low-energy weak-charge and parity-violating-asymmetry measurements, interpreted in the same scheme with the required radiative corrections~\cite{Kumar:2013yoa,lowEnergysw2}.  Together, they test the running of the weak mixing angle over a wider range of scales.

\FloatBarrier

Hadron-collider measurements provide determinations of $\sin^2\theta^\ell_{\rm eff}$ that are experimentally independent of those from $e^+e^-$ collisions at LEP and SLD.  Their precision now makes the PDF dependence of the extraction a central issue.  Better constraints on the relevant PDF combinations are needed to improve the precision, while a common treatment of PDFs and electroweak predictions is needed for consistent comparisons between measurements.  Preserving unfolded angular distributions allows these comparisons to be updated as the predictions improve.  Complementary Z-pole observables and future high-mass measurements can extend the experimental tests of electroweak couplings and their scale dependence.

\section{Acknowledgements}
This research was supported by the U.S. Department of Energy under University of Rochester grant number DE-SC0008475, and by National Research Foundation grants RS-2024-00350406 and RS-2008-NR007227 in Korea.

\bibliographystyle{elsarticle-num}
\bibliography{MixingAngle}

\end{document}